\documentclass[aps,prd,onecolumn,12pt,superscriptaddress]{revtex4-1}
\usepackage{amssymb}
\usepackage{graphicx}
\usepackage{amsmath}
\usepackage{hyperref}
\usepackage{tensor}
\begin{document}
\title{Action growth of charged black holes with a single horizon}
\author{Rong-Gen Cai}
\email{cairg@itp.ac.cn}
\affiliation{CAS Key Laboratory of Theoretical Physics,Institute of Theoretical Physics, Chinese Academy of Sciences,Beijing 100190, China}
\affiliation{Center for Gravitational Physics, Yukawa Institute for Theoretical Physics, Kyoto University, Kyoto 606-8502, Japan}
\author{Misao Sasaki}
\email{misao@yukawa.kyoyo-u.ac.edu}
\affiliation{Center for Gravitational Physics, Yukawa Institute for Theoretical Physics, Kyoto University, Kyoto 606-8502, Japan}
\affiliation{International Research Unit of Advanced Future Studies, Kyoto University, Kyoto 606-8502, Japan}
\author{Shao-Jiang Wang}
\email{schwang@itp.ac.cn}
\affiliation{CAS Key Laboratory of Theoretical Physics,Institute of Theoretical Physics, Chinese Academy of Sciences,Beijing 100190, China}
\affiliation{Center for Gravitational Physics, Yukawa Institute for Theoretical Physics, Kyoto University, Kyoto 606-8502, Japan}
\affiliation{School of Physical Sciences, University of Chinese Academy of Sciences, No.19A Yuquan Road, Beijing 100049, China}
\date{\today}

\begin{abstract}
  According to the conjecture ``complexity equals action,'' the complexity of a holographic state is equal to the action of a Wheeler-DeWitt (WDW) patch of black holes in anti-de Sitter space. In this paper we calculate the action growth of charged black holes with a single horizon, paying attention to the contribution from a spacelike singularity inside the horizon. We consider two kinds of such charged black holes: one is a charged dilaton black hole, and the other is a Born-Infeld black hole with $\beta^2 Q^2<1/4$. In both cases, although an electric charge appears in the black hole solutions, the inner horizon is absent; instead a spacelike singularity appears inside the horizon.  We find that the action growth of the WDW patch of the charged black hole is finite and satisfies the Lloyd bound. As a check, we also calculate the action growth of a charged black hole with a phantom Maxwell field. In this case, although the contributions from the bulk integral and the spacelike singularity are individually divergent, these two divergences just cancel each other and a finite action growth is obtained. But in this case, the Lloyd bound is violated as expected.
\end{abstract}
\begin{flushright}
  YITP-17-13
\end{flushright}
\maketitle

\section{Introduction}

The AdS/CFT correspondence says that quantum gravity in anti-de Sitter (AdS) space is dual to conformal field theory (CFT) living on the boundary of the AdS space~\cite{Maldacena:1997re,Gubser:1998bc,Witten:1998qj}. It is a beautiful realization of the holography of gravity~\cite{tHooft:1993dmi,Susskind:1994vu}. Thanks to the weak/strong duality in the AdS/CFT correspondence, one is able to get various properties of a strong coupling CFT by studying a weak coupling gravity theory in the bulk. Indeed, various applications of the correspondence in low energy QCD, quark gluon plasma, hydrodynamics, condensed matter theory, and even optics have appeared in recent years, and remarkable progress has been made. Very recently, Brown {\it et al.} made a conjecture that ``complexity=action''~\cite{Brown:2015bva,Brown:2015lvg}, which says that the quantum complexity of a holographic state is given by the classical action of the ``Wheeler-DeWitt (WDW) patch'' in the bulk. The quantum complexity is the minimal number of elemental operators (quantum gates) needed to produce a state of interest from a reference state. Since the action of a WDW patch in the AdS bulk is divergent, Brown {\it et al.} calculated the action growth of the WDW patch at late times. According to the ``complexity=action'' conjecture, the action growth should be equal to the complexity growth of a holographic state. The rate of the growth of complexity is just the rate of the growth of the number of simple gates needed to prepare the state of interest from a reference state. They showed that the complexity growth for a holographic state dual to an AdS Schwarzschild black hole saturates the Lloyd bound~\cite{Lloyd}.

For an eternal AdS Schwarzschild black hole, there are two boundaries denoted by times $t_L$ and $t_R$ on the left and right AdS boundaries, respectively. The black hole determines a state $|\psi (t_L,t_R)\rangle$,
\begin{equation}
 |\psi(t_L,t_R)\rangle=\mathrm{e}^{-i(H_Lt_L+H_Rt_R)}|\mathrm{TFD}\rangle,
\end{equation}
where $H_L$ and $H_R$ are the Hamiltonians on the left and right boundaries, respectively, and $|\mathrm{TFD}\rangle $ is the thermofield double state,
\begin{equation}
 |\mathrm{TFD}\rangle=Z^{-1/2}\sum_n\mathrm{e}^{-\beta E_n/2}|E_n\rangle_L\times|E_n\rangle_R.
\end{equation}
A thermofield double (TFD) state is a maximally entangled state with the reduced density matrix on either side being a usual thermal state. The conjecture in~\cite{Brown:2015bva,Brown:2015lvg} means that the complexity of the state $|\psi(t_L,t_R)\rangle$ is given by
\begin{equation}
 \label{eq3}
 {\cal C}(|\psi(t_L,t_R)\rangle=\frac{\cal A}{\pi \hbar},
\end{equation}
where ${\cal A}$ is the bulk classical action of the  WDW patch. It was found that at late times, the action growth for the AdS Schwarzschild black hole reads
\begin{equation}
 \label{eq4}
 \frac{\mathrm{d}{\cal A}}{\mathrm{d}t}=2M,
\end{equation}
where $M$ is the mass of the black hole and $t=t_L+t_R$. This implies the AdS Schwarzschild black hole saturates the Lloyd bound~\cite{Lloyd} for the state $|\psi(t_L,t_R)\rangle$. Furthermore, Brown {\it et al.}~\cite{Brown:2015bva,Brown:2015lvg} suggested an upper bound for the complexity growth of a state with certain conserved charges. By explicit calculations for the rotating Banados-Teitelboim-Zanelli (BTZ) black hole and AdS Reissner-Nordstr\"om (AdS-RN) black hole, they showed that rotating BTZ black holes and small AdS-RN black holes saturate the upper bound, while intermediate and large AdS-RN black holes violate the upper bound. One of the reasons they argued is that for small black holes it is due to the Bogomol'nyi-Prasad-Sommerfield (BPS) bound in supersymmetric theory, while in the case of intermediate and large charged black holes, the AdS-RN black holes are not a proper dual description of an UV-complete holographic field theory.

In a previous paper~\cite{Cai:2016xho}, we calculated the action growth of the WDW patch in the cases of AdS-RN black holes, (charged) rotating BTZ black holes, AdS Kerr black holes, and (charged) Gauss-Bonnet black holes. It was found that even for small charged black holes, the upper bound proposed by Brown {\it et al.}~\cite{Brown:2015bva,Brown:2015lvg} is also violated, and based on our calculations, we further suggested an expression for the action growth
\begin{equation}
 \label{eq5}
 \frac{{\mathrm{d}\cal A}}{\mathrm{d}t}=(M-\Omega J-\mu Q)_+ - (M-\Omega J-\mu Q)_-,
\end{equation}
where $\Omega$ and $\mu$ are angular velocity and chemical potential of a black hole, while $J$ and $Q$ are the angular momentum and electric charge of the black hole, respectively, and the subscripts ``$+$'' and``$-$'' stand for the angular velocity and chemical potential evaluated at the outer and inner horizons of the black hole, respectively. The upper bound (\ref{eq5}) should be saturated for stationary AdS black holes. The results of~\cite{Cai:2016xho} were further checked in the case of massive gravity~\cite{Pan:2016ecg} and in the more general cases~\cite{Huang:2016fks} where it has been proved that the action growth rate can be expressed as the difference of the generalized enthalpy between the two corresponding horizons. It has been proved in~\cite{Yang:2016awy} that under the strong energy condition of steady matter outside the Killing horizon, the action growth rate of black holes obeys the Lloyd bound~\cite{Lloyd}. It has also been argued in~\cite{Brown:2017jil} that there exists ``the second law of quantum complexity'' that the complexity growth rate of the quantum system parallels the growth of entropy of the classical system.

The so-called WDW patch (see Fig. 1 in~\cite{Brown:2015bva}) is defined as the bulk domain of dependence of a Cauchy slice anchored as the boundary state at the times $t_L$ and $t_R$ of an eternal AdS black hole. Namely, it is the spacetime region sandwiched between forward and backward light rays sending from the boundaries at $t_L$ and $t_R$, respectively.  At late times,  if there appears an inner horizon inside the black hole, the forward light rays terminate there. Otherwise, they end at the singularity. When the time at, say, the left boundary has a shift from $t_L$ to $t_L+\delta$, the change in the boundaries of the WDW patch consists of some null segments and codimension two surfaces at which a null segment is jointed to another (spacelike, timelike, or null) segment. Therefore the action growth of the WDW patch has contributions from the null segments and the codimension two surfaces. The calculations made in~\cite{ Brown:2015bva,Brown:2015lvg}  have been questioned recently by Lehner {\it et al.} in~\cite{Lehner:2016vdi}, because some ambiguities appear in the contributions from a null segment, its contribution depends on an arbitrary choice of parametrization for the generators, and similar ambiguities also appear in the contribution from those codimension two surfaces.

By a detailed analysis, Lehner {\it et al.}~\cite{Lehner:2016vdi} computed the time rate of the bulk action for the WDW patch of a black hole in AdS space.  The ambiguity from the null segments is tamed by insisting that the null generators are affinely parametrized so that the contribution from each null segment vanishes and the freedom to rescale the affine parameter by a constant factor on each generator remains unchanged.  The ambiguity in the joint contributions can also be eliminated by formulating well-motivated rules which ensure the additivity of the gravitational action.  It turns out that the two approaches, one proposed by Brown {\it et al.}~\cite{ Brown:2015bva,Brown:2015lvg} and the other by Lehner {\it et al.}~\cite{Lehner:2016vdi}, give the same results for the AdS Schwarzschild and AdS-RN black holes, although these two approaches are totally different. The paper~\cite{Lehner:2016vdi} gives a detailed comparison between the two approaches and argues possible reasons why they give the same results. Clearly these two approaches are not equivalent, and it is interesting to see in which instance they will always give the same results.

In this paper we will calculate the action growth of a WDW patch for two kinds of charged  black holes in asymptotically AdS space.  One is a charged dilaton black hole where a dilaton field appears, and the other is a Born-Infeld black hole. The motivation to perform such a calculation is twofold.  The first is to see what happens when a scalar curvature singularity  replaces the inner horizon of a charged black hole.  As we know,  when a charge appears, as in the case of AdS-RN  black holes, an inner (Cauchy) horizon may appear. But the Cauchy horizon is perturbatively unstable; hence it will turn into a curvature singularity. The second is to notice the fact that for the AdS Schwarzschild black hole case, both approaches mentioned above involve the surface term contribution from the spacelike singularity at the origin $r=0$, but the divergence in the extrinsic curvature  of the surface is remarkably canceled by the determinant of the spacelike surface in the limit $r=0$. We want to know whether such a cancellation can always happen in the other cases of spacelike singularities.  Finally as a by-product, we want to see whether the two approaches can give the same results in the examples we will discuss. The charged dilaton black hole and Born-Infled black hole in AdS space will serve for our aim in this paper. For these two kinds of black holes, the inner horizon disappears, and instead a spacelike singularity appears. Here we should mention that for the Born-Infeld black hole, the Born-Infeld coefficient $\beta$ and the charge of the black hole have to satisfy the condition $\beta^2 Q^2<1/4$; otherwise the inner horizon will appear. As a check, we also calculate the action growth of charged black holes with a phantom Maxwell field. In this case, the inner horizon is also absent.

The organization of the paper is as follows. In Sec.\ref{dilaton} we will calculate the action growth of the WDW patch in the charged dilaton black hole in asymmetrically AdS space. We will discuss the case of the Born-Infeld black hole in Sec.~\ref{BI}. The charged black hole with a phantom Maxwell field will also be discussed there. The conclusions and discussions will be presented in Sec.~\ref{Con}.
Throughout this paper we set the Newtonian constant $G=1$.

\section{Charged dilaton black hole in AdS space}
\label{dilaton}

In this section we consider the charged dilaton black hole solution in the Einstein-Maxwell-dilaton theory with action~\cite{Gao:2004tu}
\begin{equation}
 \label{eq6}
 S=\frac{1}{16\pi}\int\mathrm{d}^4x\sqrt{-g}(R-2(\partial\phi)^2-V(\phi)-\mathrm{e}^{-2\phi}F^2),
\end{equation}
where $F$ denotes the Maxwell field strength, and the potential $V(\phi)$ of the dilaton field is
given by
\begin{equation}
 V(\phi)=-\frac{4}{l^2}-\frac{1}{l^2}\left[\mathrm{e}^{2(\phi-\phi_0)}+\mathrm{e}^{-2(\phi-\phi_0)}\right],
\end{equation}
where $\phi_0$ is a constant and $l$ is the AdS radius. When $\phi=\phi_0$, the potential reduces to a negative cosmological constant with $V=-6/l^2$. Varying the action~\eqref{eq6}, one has the equations of motion,
\begin{align}
\label{eq8}
 &R_{\mu\nu}=2\partial_{\mu}\phi\partial_{\nu}\phi+\frac{1}{2}g_{\mu\nu}V+2\mathrm{e}^{-2\phi}(F_{\mu\alpha}F_{\nu}^{\,\alpha}-\frac{1}{4}g_{\mu\nu} F^2);\\
 &\partial_{\mu}(\sqrt{-g}\mathrm{e}^{-2\phi}F^{\mu\nu})=0;\\
 &\partial^2\phi=\frac{1}{4}\frac{\mathrm{d}V}{\mathrm{d}\phi}-\frac{1}{2}\mathrm{e}^{-2\phi}F^2.
\end{align}
The theory has a static spherically symmetric charged dilaton black hole solution~\cite{Gao:2004tu}
\begin{align}
 \label{eq11}
 \mathrm{d}s^2&=-f(r)\mathrm{d}t^2+f^{-1}(r)\mathrm{d}r^2+U^2(r)\mathrm{d}\Omega^2;\\
 F_{tr}&=\frac{Q\mathrm{e}^{2\phi}}{U^2},\quad\mathrm{e}^{2\phi}=\mathrm{e}^{2\phi_0}\left(1-\frac{2D}{r}\right),
\end{align}
where
\begin{align}
 f(r)&=1-\frac{2M}{r}+\frac{r(r-2D)}{l^2},\\
 \label{eq14}
 U^2(r)&=r(r-2D),\quad D=\frac{Q^2\mathrm{e}^{2\phi_0}}{2M},
\end{align}
with $M$ and $Q$ as the mass and charge of the black hole, respectively. The black hole has only one horizon $r_+$ determined by the equation $f(r)|_{r=r_+}=0$. And it is easy to see that there are two singularities inside the horizon: one is at $r=0$, and the other is at $r=2D$. In other words, there is no inner horizon for this charged black hole solution. The Penrose diagram of the black hole is quite different from the one for an AdS-RN black hole, but similar to the one for an AdS Schwarzschild black hole. In this case, the spacetime ends at $r=2D$ since the region $0<r<2D$ is completely disconnected from the outer spacetime. Hence the light rays of a WDW patch will end at the singularity at $r=2D$, rather than at $r=0$.

\begin{figure}
\includegraphics[width=0.55\textwidth]{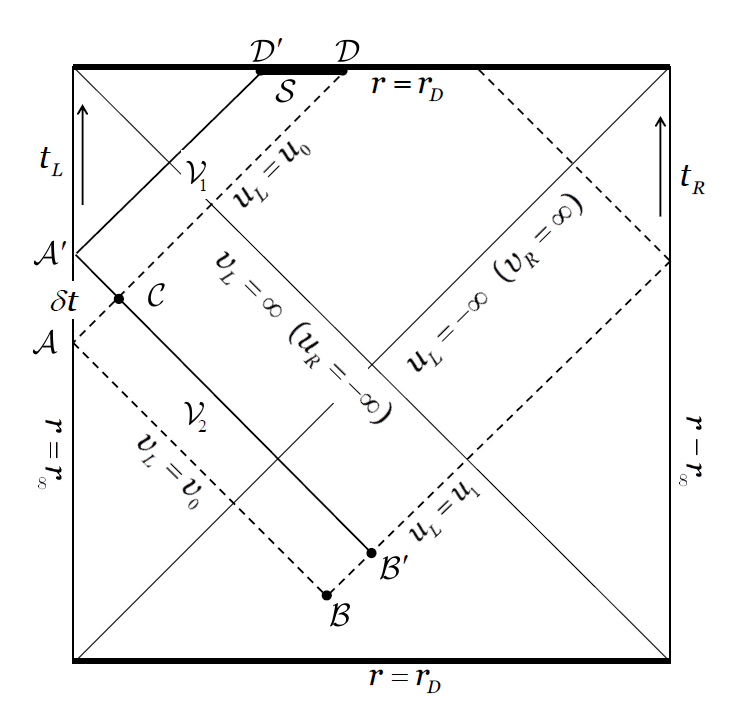}%
\includegraphics[width=0.4\textwidth]{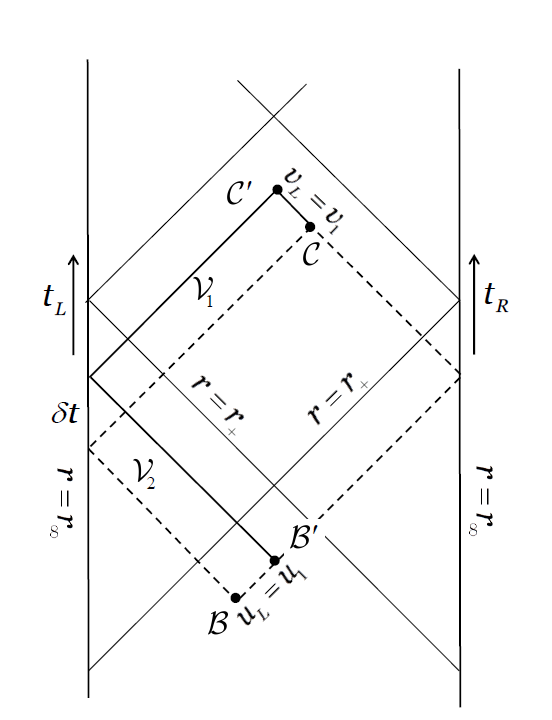}\\
  \caption{A WDW patch and its change due to an infinitesimal time shift $\delta t$ at the left boundary, either for a charged dilaton AdS black hole (BH) or for a Born-Infeld (BI) AdS BH with single horizon (left panel), and the same for a BI-AdS BH with two horizons (right panel). For the charged dilaton AdS BH, the light rays from the boundaries terminate on a nonzero spacelike singularity at $r=r_D=2D$, while for the BI-AdS BH with single horizon, they terminate at $r=r_D=0$. For the BI-AdS BH with two horizons the light rays from both boundaries meet without encountering a singularity. Figures are taken from~\cite{Lehner:2016vdi} and slightly modified.}\label{fig:WDWpatch}
\end{figure}

Now we employ the approach in~\cite{Lehner:2016vdi} to calculate the action change, $\delta S=S(t_0+\delta t)-S(t_0)$, of the WDW patch when the time has a shift from $t_0$ to $t_0 +\delta t$, say on the left boundary. See the left panel of Fig.~\ref{fig:WDWpatch},
which is essentially a reproduction of Fig.~12 in~\cite{Lehner:2016vdi}. Note that in this approach, each null surface has no contribution to the action due to an affine parametrization. The null joint at $r=r_{\infty}$ has a contribution to the action $S(t_0+\delta t)$, which is simply equal to the contribution to the action $S(t_0)$ due to the time translation. The same holds for the joints linking the incoming null segment to the spacelike surface near the singularity at $r=2D$. This means that both joints have no contribution to the action change $\delta S$. As a result, the action change comes from the volume contributions from the regions ${\cal V}_1$ and ${\cal V}_2$, the surface contribution from the spacelike segment ${\cal S}$, and the joint contributions from the surfaces at ${\cal B'}$ and ${\cal B}$. Namely we have
\begin{equation}
 \label{eq15}
 \delta S=S_{{\cal V}_1}-S_{{\cal V}_2}-\frac{1}{8\pi}\int_{\cal S}K\mathrm{d}\Sigma+\frac{1}{8\pi}\int_{\cal B'}\,a\mathrm{d}S-\frac{1}{8\pi}\int _{\cal B}\,a\mathrm{d}S,
\end{equation}
where $\mathrm{d}\Sigma$ is the volume element on ${\cal S}$, $\mathrm{d}S$ is the surface element on ${\cal B}'$ and ${\cal  B}$, and the integrand $a$ is given later by Eq.~(\ref{eq22}).

As in~\cite{Lehner:2016vdi}, we introduce two null coordinates $u_L$ and $v_L$ by $u_{L}=t_{L}+r^*$ and $v_{L}=t_{L}-r^*$ where $r^*=\int f^{-1}\mathrm{d}r$, where $t_L$ is the future-directed time on the left boundary. Note that one may similarly introduce null coordinates $u_R$ and $v_R$ with respect to the future-directed time in the right region. Since $t_L$ and $t_R$ are related analytically by $t_R=-t_L+\mathrm{const}.$, it follows that $v_R=-u_L+\mathrm{const}.$ and $u_R=-v_L+\mathrm{const}.$ in the sense of analytic continuation. Below we focus on the coordinates defined in the left region, and we omit the index $L$ from the null coordinates unless there may arise any confusion.

As shown in the left panel for Fig~\ref{fig:WDWpatch}, the past and future null boundaries on the left of the first WDW patch are labeled by
$u=u_0$ and $v=v_0$, respectively. These null boundaries become $u=u_0+\delta t$ and $v=v_0+\delta t$ after a time shift $\delta t$. Thus the region ${\cal V}_1$ is surrounded by the null surfaces $u=u_0$, $u=u_0+\delta t$, $v=v_0+\delta t$, and the spacelike surface ${\cal S}$ at $r=r_D+\epsilon$ near the singularity $r_D\equiv2D$. At the end of calculations, we will take the limit $\epsilon\rightarrow0$. The surface $v=v_0+\delta t$ is described by $r=\rho(u)$, where $\rho(u)$ is defined implicitly by $r^*(\rho)=-\frac{1}{2}(v_0+\delta t-u)$. Using Eq.~\eqref{eq8}, thus we have
\begin{align}
 16\pi S_{{\cal V}_1}&=-\frac{4\pi}{l^2}\int_{u_0}^{u_0+\delta t}\mathrm{d}u\int_{r_D+\epsilon}^{\rho(u)}\mathrm{d}r\left(4r(r-2D)+(r-2D)^2+r^2-\frac{2Q^2l^2\mathrm{e}^{2\phi_0}}{r^2}\right)\nonumber\\
 \label{eq16}
 &=-\frac{4\pi}{l^2}\int_{u_0}^{u_0+\delta t}\mathrm{d}u\left[2\rho(u)^3-6D\rho(u)^2+4D^2\rho(u)+\frac{2Q^2l^2\mathrm{e}^{2\phi_0}}{\rho(u)}\right]+\frac{4\pi Q^2 \mathrm{e}^{2\phi_0}}{D}\delta t.
\end{align}
In the second line, we have already taken the limit $\epsilon\rightarrow0$. Note that the above volume integral has no divergence from the singularity at $r=2D$.

As shown in the left panel of Fig.~\ref{fig:WDWpatch}, the region ${\cal V}_2$ is bounded by the null surfaces $u=u_0$, $u=u_1$, $v=v_0$, and $v=v_0+\delta t $. In this case, the volume integral in ${\cal V}_2$ should be done in the $(v,r)$ coordinates, where the surfaces $u=u_{0/1}$ are described by $r=\rho_{0/1}(v)$ with $r^*(\rho_{0/1})=-\frac{1}{2}(v-u_{0/1})$. The result turns out to be
\begin{align}
 16\pi S_{{\cal V}_2}&=-\frac{4\pi}{l^2}\int_{v_0}^{v_0+\delta t}\mathrm{d}v\int_{\rho_1(v)}^{\rho_0(v)}\mathrm{d}r\left(4r(r-2D)+(r-2D)^2+r^2-\frac{2Q^2l^2\mathrm{e}^{2\phi_0}}{r^2}\right)\nonumber \\
 \label{eq17}
 &=\left.-\frac{4\pi}{l^2}\int_{v_0}^{v_0+\delta t}\mathrm{d}v\left(2r^3-6Dr^2+4D^2r+\frac{2Q^2l^2\mathrm{e}^{2\phi_0}}{r}\right)  \right|_{r=\rho_1(v)}^{r=\rho_0(v)}.
\end{align}
After making a change of the variable $u\rightarrow u_0+v_0+\delta t-v$ in Eq.~\eqref{eq16}, the integration bound $[u_0,u_0+\delta t]$ becomes $[v_0+\delta t,v_0]$, and $\rho(u)$ which satisfies $r^*(\rho)=-\frac12(v_0+\delta t-u)\rightarrow-\frac12(v-u_0)$ becomes $\rho_0(v)$. Then combining Eq.~\eqref{eq16} with Eq.~\eqref{eq17}, we have
\begin{equation}
 \label{eq18}
 16\pi (S_{{\cal V}_1} -S_{{\cal V}_2})= \frac{4\pi Q^2  e^{2\phi_0}}{D} \delta t -\frac{4\pi}{l^2} \int^{v_0+\delta t}_{v_0} dv \left (2\rho^3_1-6D\rho^2_1 +4D^2\rho_1 +\frac{2Q^2 l^2e^{2\phi_0}}{\rho_1}\right).
\end{equation}
The variable $\rho_1(v)$ varies from $r_{\cal B}$ to $r_{\cal B'}$ as $v$ increases from $v_0$ to $v_0+\delta t$.  But the variation is small, and one has $r_{\cal B'}=r_{\cal B}+{\cal O}(\delta t)$ so that Eq.~\eqref{eq18} reduces to
\begin{equation}
 \label{eq19}
 16\pi(S_{{\cal V}_1}-S_{{\cal V}_2})=\frac{4\pi Q^2\mathrm{e}^{2\phi_0}}{D}\delta t-\frac{4\pi}{l^2}\left(2r_{\cal B}^3-6Dr_{\cal B}^2+4D^2r_{\cal B}+\frac{2Q^2l^2\mathrm{e}^{2\phi_0}}{r_{\cal B}}\right)\delta t.
\end{equation}

Next we calculate the contribution from the spacelike surface ${\cal S}$ at $r=2D+\epsilon$. Its unit normal vector is given by $n_{\alpha}=|f|^{-1/2}\partial_{\alpha}r$, and the extrinsic curvature reads
\begin{equation}
 K=\nabla_{\alpha}n^{\alpha}=\frac{1}{U^2}\frac{\mathrm{d}}{\mathrm{d}r}\left(U^2|f|^{1/2}\right),
\end{equation}
and the surface element $\mathrm{d}\Sigma=4\pi|f|^{1/2}U^2\mathrm{d}t$. Thus the contribution from the spacelike surface has the form
\begin{equation}
 \label{eq21}
 16\pi S_K=-2\int_{\cal S}K\mathrm{d}\Sigma=16\pi(M-D)\delta t,
\end{equation}
where we have considered the condition with $M>D$.

Finally let us calculate the contributions from the joints ${\cal B'}$ and ${\cal B}$.  Following Ref.~\cite{Lehner:2016vdi}, the integrand $a$ in Eq.~\eqref{eq15} has the form
\begin{equation}
 \label{eq22}
 a=\ln(-\frac{1}{2}k\cdot\bar k),
 \end{equation}
where $k^{\alpha}$ is the future-directed null normal to the left-moving null surface, i.e., on which $v=v_0$ and $v=v_0+\delta t$, while $\bar k^{\alpha}$ is future-directed null normal to the right-moving surface, on which $u= u_1$. Under affine parametrization, they can be expressed as
\begin{equation}
 \label{eq23}
 k_{\alpha}=-c\partial_{\alpha}v=-c\partial_{\alpha}(t-r^*), \quad \bar k_{\alpha}=\bar c\partial_{\alpha}u=\bar c\partial_{\alpha}(t+r^*),
\end{equation}
where $c$ and $\bar c$ are two arbitrary positive constants. This implies that one has the asymptotic normalizations $k\cdot \hat{t}_L=-c$ and $\bar{k}\cdot\hat{t}_R=-\bar c$, where $\hat{t}_{L,R}$ are the asymptotic Killing vectors which are normalized to describe the time flow in the left and right boundary theories, respectively. With the choice, one has $a=-\ln(-f/c\bar c)$,  and
\begin{equation}
 16\pi S_{\cal B'B}=2\int_{\cal B'}a\mathrm{d}S-2\int_{\cal B}a\mathrm{d}S=8\pi[h(r_{\cal B'})-h(r_{\cal B})],
\end{equation}
where $h(r)=-U^2\ln(-f/c\bar c)$. Making a Taylor expansion of $h(r)$ around $r=r_{\cal B}$ and using $r_{\cal B'}-r_{\cal B}=-\frac{1}{2}f\delta t$,  one obtains
\begin{equation}
 \label{eq25}
 16\pi S_{\cal B'B}=\left.4\pi\delta t\left[U^2\frac{\mathrm{d}f}{\mathrm{d}r}+2(r-D)f\ln\left(\frac{-f}{c\bar c}\right)\right]\right|_{r=r_{\cal B}}.
\end{equation}
Putting~\eqref{eq19},~\eqref{eq21}, and~\eqref{eq25} together, we have
\begin{align}
 16\pi\frac{\mathrm{d}S}{\mathrm{d}t}&=\frac{4\pi Q^2\mathrm{e}^{2\phi_0}}{D}-\frac{4\pi}{l^2}\left(2r_{\cal B}^3-6Dr_{\cal B}^2+4D^2r_{\cal B}+\frac{2Q^2l^2\mathrm{e}^{2\phi_0}}{r_{\cal B}}\right)\nonumber\\
&+16\pi(M-D)+\left.4\pi\left[U^2\frac{\mathrm{d}f}{\mathrm{d}r}+2(r-D)f\ln\left(\frac{-f}{c\bar c}\right)\right]\right|_{r=r_{\cal B}}.
\end{align}
At late times, $r_{\cal B}$ approaches $r_+$, and $f(r)$ goes to zero. As a result, we finally arrive at
\begin{align}
 \frac{\mathrm{d}S}{\mathrm{d}t}
&=2M-Q^2\mathrm{e}^{2\phi_0}\left(\frac{1}{2M}+\frac{1}{r_+}\right)
\nonumber\\
&=2M-\mu_+Q-D,\quad D=\frac{Q^2\mathrm{e}^{2\phi_0}}{2M},\label{eq:dilation}
\end{align}
where the chemical potential $\mu_+=Q^2\mathrm{e}^{2\phi_0}/r$ is used. When $Q=0$, the dilaton black hole solution~\eqref{eq11}-\eqref{eq14} reduces to the AdS Schwarzschild solution, while the action growth~\eqref{eq:dilation} goes back to the one for the AdS Schwarzschild black hole,~\eqref{eq4}, as expected. Of course, the action growth~\eqref{eq:dilation} is clearly different from the one
for the AdS-RN black hole.

As a comparison, we now naively follow the approach proposed by Brown {\it et al.}~\cite{Brown:2015bva,Brown:2015lvg} to present the corresponding result for the charged dilaton black hole discussed above. In this case, the contribution from the bulk term is given by
\begin{align}
16\pi S_{\rm bulk}&=4\pi\delta t\int_{r_D}^{r_+}\mathrm{d}r\left(-\frac{4r(r-2D)}{l^2}-\frac{1}{l^2}((r-2D)^2 +r^2)+\frac{2Q^2\mathrm{e}^{2\phi_0}}{r^2}\right)\nonumber\\
\label{eq29}
&=8\pi\delta t\left[-\frac{1}{l^2}(r_+^3-3Dr_+^2+2D^2r_+)+Q^2\mathrm{e}^{2\phi_0}\left(\frac{1}{r_D}-\frac{1}{r_+}\right)\right],
\end{align}
while the surface term contributes
\begin{align}
16\pi S_{\rm surf}&
=\left.8\pi\delta tf^{1/2}\frac{\mathrm{d}(U^2f^{1/2})}{\mathrm{d}r}\right|^{r_+}_{r_D}
\nonumber\\
\label{eq30}
&=8\pi\delta t\left(3M-2D+\frac{1}{l^2}(r_+^3-3Dr_+^2+2D^2r_+)-\frac{Q^2\mathrm{e}^{2\phi_0}}{r_+}\right).
\end{align}
Combining~\eqref{eq29} and~\eqref{eq30}, we have
\begin{equation}
\label{eq31}
\frac{\mathrm{d}S}{\mathrm{d}t}=2M-Q^2\mathrm{e}^{2\phi_0}\left(\frac{1}{2M}+\frac{1}{r_+}\right).
\end{equation}
Clearly when $Q=0$, the above result does reduce to the one~\eqref{eq4} for the AdS Schwarzschild black hole. Further it is interesting to note that Eq.~\eqref{eq31} exactly matches the one we obtained in~\eqref{eq:dilation}. Although these two approaches are totally different, we see that they produce the same result in this example.

\section{Born-Infeld black hole in AdS space}
\label{BI}

In this section, we consider another example of charged black holes in AdS space, the so-called Born-Infeld black hole. The theory is described by
the following action:
\begin{equation}
\label{eq32}
S=\frac{1}{16\pi}\int\mathrm{d}^4x\sqrt{-g}[R+\frac{6}{l^2}+{\cal L}(F)],
\end{equation}
where
\begin{equation}
{\cal L}(F)=4\beta^2\left(1-\sqrt{1+\frac{F_{\mu\nu}F^{\mu\nu}}{2\beta^2}}\right).
\end{equation}
When $\beta\rightarrow\infty$, the Born-Infeld theory reduces to the Maxwell theory with ${\cal L}(F)=-F^2$. The Einstein field equations from the
action~\eqref{eq32} read
\begin{equation}
 \label{eq34}
 R_{\mu\nu}-\frac{1}{2}g_{\mu\nu}R-\frac{3}{l^2}g_{\mu\nu}=\frac{1}{2}g_{\mu\nu}{\cal L}(F)+\frac{2F_{\mu\alpha}F_{\nu}^{\,\alpha}}{\sqrt{1+F^2/(2\beta^2)}}.
\end{equation}
The equations of motion admit the static symmetrically symmetric black hole solution~\cite{Rasheed:1997ns,Fernando:2003tz,Dey:2004yt,Cai:2004eh}
\begin{align}
\label{eq35}
 \mathrm{d}s^2&=-f(r)\mathrm{d}t^2+f^{-1}(r)\mathrm{d}r^2+r^2\mathrm{d}\Omega^2;\\
 F_{tr}&=\frac{Q}{\sqrt{r^4+Q^2/\beta^2}},
\end{align}
where
\begin{align}
f(r)&=1-\frac{2M}{r}+\frac{r^2}{l^2}+\frac{2\beta^2}{r}\int_r^\infty\mathrm{d}x\left(\sqrt{x^4+Q^2/\beta^2}-x^2\right)
\nonumber\\
&=1-\frac{2M}{r}+\frac{r^2}{l^2}+\frac{2\beta^2}{r}\left[I(r=\infty;a^2=\frac{Q^2}{\beta^2})-I(r;a^2=\frac{Q^2}{\beta^2})\right],
\label{eq38}
\end{align}
where $I(r;a^2)\equiv\int^r\mathrm{d}x\left(\sqrt{x^4+a^2}-x^2\right)$ is introduced for the simplicity of later computations. In the solution, $M$ and $Q$ are the mass and charge of the Born-Infeld black hole, respectively. For $r\rightarrow\infty$, $I(r;a^2=\frac{Q^2}{\beta^2})$ approaches a constant $I_\infty\equiv\frac23a^\frac32\frac{\Gamma(\frac14)\Gamma(\frac54)}{\Gamma(\frac12)}$ which depends on $Q$. Therefore in the solution the term $\beta^2I_\infty$ cannot be simply absorbed into $M$. For $r^4>a^2$, $I(r;a^2)$ can be expressed as
\begin{equation}
I(r;a^2)=I_\infty-\frac{r^3}{3}+\frac{r}{3}\sqrt{r^4+a^2}-\frac{2a^2}{3r}\,_2F_1[\frac14,\frac12,\frac54,-\frac{a^2}{r^4}],
\end{equation}
which can be analytically continued into $r^4<a^2$ as
\begin{equation}
I(r;a^2)=-\frac{r^3}{3}+\frac{r}{3}\sqrt{r^4+a^2}+\frac{2}{3}r\sqrt{a^2}\,_2F_1[\frac14,\frac12,\frac54,-\frac{r^4}{a^2}].
\end{equation}
Here note that the hypergeometric function $\,_2F_1[a,b,c,z]$ is convergent for $|z|<1$. For $r\rightarrow0$, we have $I(r;a^2)\rightarrow0$.
By formally writing $I(r;a^2)=-\frac{r^3}{3}+\frac{r}{3}\sqrt{r^4+a^2}+\frac{2}{3}F(r;a^2)$, where $F'(r;a^2)=a^2/\sqrt{r^4+a^2}$, the chemical potential can be calculated as
\begin{align}
\mu(r)Q&=\int_r^\infty\frac{Q^2\mathrm{d}x}{\sqrt{x^4+Q^2/\beta^2}}
=\beta^2\left.F(r;\frac{Q^2}{\beta^2})\right|_r^\infty
\nonumber\\
  &=\frac{Q^2}{r}\,_2F_1[\frac14,\frac12,\frac54,-\frac{Q^2}{\beta^2r^4}],
\end{align}
where in the second line we have used the relation obtained from the analytic continuation,
\begin{equation}
F(r;a^2)=\frac{3}{2}I_\infty
-\frac{a^2}{r}\,_2F_1[\frac14,\frac12,\frac54,-\frac{a^2}{r^4}].
\end{equation}

In a generic case, the Born-Infeld black hole~\eqref{eq35}-\eqref{eq38} has two horizons, an outer horizon $r_+$ and an inner horizon $r_-$. Both of them satisfy the equation $f(r)=0$. The outer horizon $r_+$ is the large positive root of the equation $f(r)|_{r=r_+}=0$. And the associated temperature is given by
\begin{equation}
T=\frac{1}{4\pi r_+}\left[1+\frac{3r_+^2}{l^2}+2\beta^2r_+^2\left(1-\sqrt{1+\frac{Q^2}{\beta^2r_+^4}}\right)\right].
\end{equation}
An extremal black hole corresponds to a zero temperature black hole with $T=0$. Solving the equation $T=0$, one has
\begin{equation}
r_+^2=\frac{l^2}{6}\frac{1+\frac{3}{2\beta^2l^2}}{1+\frac{3}{4\beta^2l^2}}\left(-1\pm\sqrt{1+\left(\beta^2Q^2-\frac{1}{4}\right) \frac{12\left(1+\frac{3}{4\beta^2 l^2}\right)}{\beta^2l^2\left(1+\frac{3}{2\beta^2l^2}\right)^2}}\right).
\end{equation}
We see that for the branch with sign ``$-$'' there is no real positive root, while for the branch with sign `` $+$," only when $\beta^2Q^2\ge1/4$, one has a real positive root. This indicates that only when $\beta^2Q^2\ge1/4$ can the black hole solution have two horizons, and an extremal black hole can be achieved, while when $\beta^2Q^2<1/4$, the black hole has only one horizon and the inner horizon is absent. In other words, when $\beta^2Q^2\ge1/4$, the Penrose diagram of the Born-Infeld black hole in AdS space is similar to the one for the AdS-RN black hole, while when $\beta^2Q^2<1/4$, the Penrose diagram is similar to the one for the AdS Schwarzschild black hole, although the Born-Infeld black hole is still charged. Now we are interested in seeing what the difference of the action growth of a WDW patch  is in these two cases.

\subsection{ The case with two horizons}

In this case, the WDW patch is the same as the one for the AdS-RN black hole shown in the right panel of Fig.~\ref{fig:WDWpatch},
which is a reproduction of Fig.~13 in Ref.~\cite{Lehner:2016vdi}. Through the analysis made there, one knows that the action change has three parts, one is from the bulk integral from the inner horizon to the outer horizon, the second is from the joints ${\cal B'}$ and ${\cal B}$ inside the past horizon, and the third is from the joints ${\cal C'}$ and ${\cal C}$ inside the future horizon. The calculation can be made simply by following Ref.~\cite{Lehner:2016vdi}.

With the equations of motion~\eqref{eq34}, the Lagrangian in the bulk has the form
\begin{equation}
{\cal L}_\mathrm{bulk}=-\frac{6}{l^2}-4\beta^2+\frac{4\beta^2}{r^2}\sqrt{r^4+\frac{Q^2}{\beta^2}}.
\end{equation}
After taking the late-time limit, the bulk integral has the contribution:
\begin{align}
16\pi S_{\mathcal{V}_1-\mathcal{V}_2}
&=4\pi\delta t\int_{r_-}^{r_+}\mathrm{d}r r^2\left(-\frac{6}{l^2}-4\beta^2+\frac{4\beta^2}{r^2}\sqrt{r^4+\frac{Q^2}{\beta^2}}\right)
\nonumber\\
&=4\pi\delta t\left.\left[-\frac{2}{l^2}r^3+4\beta^2I(r;\frac{Q^2}{\beta^2})\right]\right|_{r_-}^{r_+},
\label{eq42}
\end{align}
while the joints have the following contributions,
\begin{align}
16\pi S_{\mathcal{B'B}-\mathcal{C'C}}
&=4\pi\delta t\left.r^2f'(r)\right|_{r_-}^{r_+}
\nonumber\\
\label{eq43}
&=4\pi\delta t\left.\left[2M+\frac{2r^3}{l^2}-2\beta^2(I_\infty-I_r)+2\beta^2r^3-2\beta^2r\sqrt{r^4+\frac{Q^2}{\beta^2}}\right]\right|_{r_-}^{r_+}.
\end{align}
Combining~\eqref{eq42} and~\eqref{eq43}, we obtain
\begin{align}
\frac{\mathrm{d}S}{\mathrm{d}t}
&=\left.\left[\frac{3}{2}\beta^2I_r+\frac{\beta^2}{2}r^3-\frac{\beta^2}{2}r\sqrt{r^4+\frac{Q^2}{\beta^2}}\right]\right|_{r_-}^{r_+};\\
\label{eq44}
&=\left.\left(\frac{Q^2}{r}\,_2F_1[\frac14,\frac12,\frac54,-\frac{Q^2}{\beta^2r^4}]\right)\right|_{r_+}^{r_-};\\
&\equiv\mu_-Q-\mu_+Q.
\end{align}
Here $\mu_+$ and $\mu_-$ are chemical potentials associated with the charge $Q$ of the black hole at the outer horizon and inner horizon~\cite{Cai:2004eh}, respectively.  When $\beta^2\rightarrow\infty$, the hypergeometric function $\,_2F_1[\frac14,\frac12,\frac54,-\frac{Q^2}{\beta^2r^4}]$ in~\eqref{eq44} reduces to unity. As expected, our result here~\eqref{eq44} reduces to the one for the AdS-RN black hole~\cite{Brown:2015bva,Brown:2015lvg,Cai:2016xho,Lehner:2016vdi} . Furthermore, let us notice that the result~\eqref{eq44} can be further expressed as
\begin{equation}
 \label{eq45}
 \frac{\mathrm{d}S}{\mathrm{d}t}=(M-\mu Q)_+-(M-\mu Q)_-.
\end{equation}
In other words, the action growth for the Born-Infeld black hole also satisfies the universal expression proposed in our previous work~\cite{Cai:2016xho}. We now move to the case with a single horizon case, which we are more interested in.

\subsection{ The case with a single horizon}

As analyzed in the above, when $\beta^2Q^2<1/4$, the Born-Infeld black hole has only one horizon; namely the inner horizon is absent. In this case, the Penrose diagram of the black hole is like the one for the AdS Schwarzschild black hole, as shown in left panel of Fig.~\ref{fig:WDWpatch}, where the origin $r=0$ is a spacelike singularity. Thus the conformal diagram is the same as the charged dilaton black hole case discussed in the previous section. Nevertheless there is an important difference that the singularity in the current case is stronger in the sense that the curvature diverges badly, while the singularity in the charged dilaton AdS is at $r=2D>0$ which is milder because it is merely a cusp singularity from the geometrical point of view. We want to see whether the contribution from the spacelike surface approaching this strong singularity is finite.

With the equations of motion~\eqref{eq34}, one is able to show that the contribution from the region ${\cal V}_1$ has the form
\begin{align}
16\pi S_{{\cal V}_1}&
=4\pi\int_{u_0}^{u_0+\delta t}\mathrm{d}u\int_{\epsilon}^{\rho(u)}\mathrm{d}r\left(-\frac{6}{l^2}r^2-4\beta^2r^2+4\beta^2\sqrt{r^4+\frac{Q^2}{\beta^2}}\right)
\nonumber\\
\label{eq47}
&=4\pi\int_{u_0}^{u_0+\delta t}\mathrm{d}u\left.\left[-\frac{2}{l}r^3+4\beta^2I(r;\frac{Q^2}{\beta})\right]\right|_\epsilon^{\rho(u)},
\end{align}
while the action in the region ${\cal V}_2$ is given by
\begin{equation}
16\pi S_{{\cal V}_2}=4\pi\int_{v_0}^{v_0+\delta t}\mathrm{d}v\left.\left[-\frac{2}{l}r^3+4\beta^2I(r;\frac{Q^2}{\beta})\right]\right|_{\rho_1(v)}^{\rho_0(v)}.
\end{equation}
Combining these two terms gives
\begin{equation}
\label{eq49}
16\pi S_{\mathcal{V}_1-\mathcal{V}_2}=4\pi\delta t\left[-\frac{2}{l^2}r_+^3+4\beta^2I_+\right].
\end{equation}

Similar to the case of the charged dilaton black hole discussed in the previous section, the contribution from the spacelike surface near the singularity is given by
\begin{align}
16\pi S_K&=-8\pi\delta t\left.\left(2rf+\frac{r^2}{2}f'\right)\right|_{r=\epsilon}
\nonumber\\
&=-8\pi\delta t\left.\left(2r-3M+\frac{3}{l^2}r^3+3\beta^2(I_\infty-I_r)+\beta^2r^3-\beta^2r\sqrt{r^4+\frac{Q^2}{\beta^2}}\right)\right|_{r=\epsilon}
\nonumber\\
&=4\pi\delta t(6M-6\beta^2I_\infty).
\label{eq53}
\end{align}
Note that the contribution from the spacelike singularity is completely finite and that there does not exist any divergence. This is the same as
the case of the AdS Schwarzschild black hole. As for the contributions from the joints ${\cal B'}$ and $ {\cal B}$, taking the same expression for $a$ in~\eqref{eq22}, we have
\begin{align}
16\pi S_{\mathcal{B'B}}&=4\pi\delta\left.r^2f'(r)\right|_{r=r_+}
\nonumber\\
&=4\pi\delta t\left[2M+\frac{2}{l^2}r_+^3-2\beta^2(I_\infty-I_+)+2\beta^2r_+^3-2\beta^2r_+\sqrt{r_+^4+\frac{Q^2}{\beta^2}}\right].
\end{align}
Putting the three contributions together, we finally arrive at
\begin{align}
\frac{\mathrm{d}S}{\mathrm{d}t}&=2M-2\beta^2I_\infty+\frac{3}{2}\beta^2I_++\frac{\beta^2}{2}r_+^3-\frac{\beta^2}{2}r_+\sqrt{r_+^4+\frac{Q^2}{\beta^2}}
\nonumber\\
\label{eq55}
&=2M-\frac{Q^2}{r_+}\,_2F_1[\frac14,\frac12,\frac54,-\frac{Q^2}{\beta^2r_+^4}]-\frac{\beta^2}{2}I_\infty.
\end{align}

First, let us note that when $Q\to 0$, this result reduces to the one for the AdS Schwarzschild black hole, as expected as a self-consistency check.
Second, similar to the form of the action growth rate for the charged dilation AdS black hole~\eqref{eq:dilation}, one can rewrite the above as
\begin{equation}
\label{eq:Born-Infeld}
\frac{\mathrm{d}S}{\mathrm{d}t}=2M -\mu_+Q-C,\quad C=\frac{\beta^2}{2}I_\infty=\beta^\frac12Q^\frac32\frac{\Gamma(\frac14)\Gamma(\frac54)}{3\Gamma(\frac12)}.
\end{equation}
The Lloyd bound is satisfied in both cases but is slowing down not only due to the presence of a Maxwell field but also due to the coupling between scalar and Maxwell fields. An interesting question is that in which case the action growth rate would speed up against the Lloyd bound. For this purpose, we next consider a model of a charged black hole with a single horizon, namely, a charged black hole with a phantom Maxwell field.

Before proceeding to it, let us first check whether the approach by Brown {\it et al.} can produce the same result as~\eqref{eq55} for the Born-Infeld AdS black hole. By using the approach in~\cite{Brown:2015bva,Brown:2015lvg}, for the Born-Infeld black hole case, we have the bulk contribution to the action as
\begin{align}
16\pi S_\mathrm{bulk}&=4\pi\delta t\int_\epsilon^{r_+}\mathrm{d}r r^2\left(-\frac{6}{l^2}-4\beta^2+\frac{4\beta^2}{r^2}\sqrt{r^4+\frac{Q^2}{\beta^2}}\right)
\nonumber\\
&=4\pi\delta t\left[-\frac{2}{l^2}r_+^3+4\beta^2I_+\right],
\end{align}
while the surface term has the form
\begin{align}
16\pi S_\mathrm{surf}
&=8\pi\delta t\left.\left(2rf+\frac{r^2}{2}f'\right)\right|_\epsilon^{r_+}
\nonumber\\
&=4\pi\delta t\left[4r_++\frac{6}{l^2}r_+^3-6\beta^2I_++2\beta^2r_+^3-2\beta^2r_+\sqrt{r_+^4+\frac{Q^2}{\beta^2}}\right].
\end{align}
Combing the above bulk and surface terms and using
$2M=r_+ + \frac{r^3_+}{l^2}+2\beta^2(I_\infty-I_+)$, we have
\begin{align}
\frac{\mathrm{d}S}{\mathrm{d}t}&=2M-2\beta^2I_\infty+\frac{3}{2}\beta^2I_++\frac{\beta^2}{2}r_+^3-\frac{\beta^2}{2}r_+\sqrt{r_+^4+\frac{Q^2}{\beta^2}}
\nonumber\\
&=2M-\frac{Q^2}{r_+}\,_2F_1[\frac14,\frac12,\frac54,-\frac{Q^2}{\beta^2r_+^4}]-\frac{\beta^2}{2}I_\infty.
\end{align}
This is exactly the same as the one in~\eqref{eq55}. Thus once again, we have shown that these two different approaches give the same result for the Born-Infeld black hole.

Finally let us mention that when $\beta^2 Q^2 \ge 1/4$, the Born-Infeld black hole will have an inner horizon $r_-$, besides the outer horizon $r_+$. In this case, the action growth is given by~\eqref{eq44}, which is of the form~\eqref{eq45} suggested in~\cite{Cai:2016xho}. But clearly it is quite different from~\eqref{eq55} for the case with a single horizon. Then a natural question arises: What is the physical meaning of the condition $\beta^2 Q^2 <1/4$ in the dual boundary field theory ? What we know from the bulk black holes is that when $\beta^2 Q^2 <1/4$, one is not able to take the limit $T\to 0$; namely like the AdS Schwarzschild black holes, there exists a minimal temperature for the dual field theory.

\subsection{Charged black hole with phantom Maxwell field}

In a generic case, an AdS-RN black hole has two horizons. But if one changes the sign of the charge term in the solution, there will be only one horizon left. Such a case can be realized within the Einstein-phantom-Maxwell theory with a negative cosmological constant,
\begin{equation}
\label{eq57}
S=\frac{1}{16\pi}\int\mathrm{d}^4x\sqrt{-g}(R+\frac{6}{l^2}+F_{\mu\nu}F^{\mu\nu}).
\end{equation}
Note that here the Maxwell term has a wrong sign. It is easy to show that this theory has the following solution:
\begin{align}
\mathrm{d}s^2&=-f(r)\mathrm{d}t^2+f^{-1}(r)\mathrm{d}r^2+r^2\mathrm{d}\Omega^2;\\
F_{tr}&=\frac{Q}{r^2};\\
f(r)&=1+\frac{r^2}{l^2}-\frac{2M}{r}-\frac{Q^2}{r^2}.
\end{align}
Here $M$ and $Q$ are the mass and charge of the solution. Clearly in this case, the solution has only one horizon $r_+$ satisfying $f(r_+)=0$,
the inner horizon in the AdS-RN black hole is absent, and the singularity at $r=0$ is spacelike. The Penrose diagram for this black hole solution is the same as the one for the AdS Schwarzschild black hole. Now we calculate the action growth of a WDW patch for this charged black hole.

By the same approach, the bulk contribution is obtained as
\begin{equation}
16\pi S_{\cal V}=-8\pi\delta t\left(\frac{r_+^3}{l^2}-\frac{Q^2}{r_+}\right)-8\pi\delta t\frac{Q^2}{\epsilon},
\end{equation}
and the contributions from the joints ${\cal B'}$ and ${\cal B}$ as
\begin{equation}
16\pi S_{\cal B'B}=8\pi\delta t\left(\frac{r_+^3}{l^2}+M+\frac{Q^2}{r_+}\right).
\end{equation}
The contribution from the spacelike singularity is given by
\begin{equation}
16\pi S_K=8\pi\delta t\left(3M+\frac{Q^2}{\epsilon}\right).
\end{equation}
Putting these three parts together we find
\begin{equation}
\label{eq64}
\frac{\mathrm{d}S}{\mathrm{d}t}=2M+\frac{Q^2}{r_+}=2M+\mu_+Q.
\end{equation}
We see that although both bulk and singularity parts are divergent as $\epsilon \to 0$, these two divergences just cancel each other in the final expression.

At this stage, let us note that both for the charged dilaton black hole and for the Born-Infeld black hole, the action (complexity) growth~\eqref{eq:dilation} and~\eqref{eq:Born-Infeld} satisfies the Lloyd bound~\eqref{eq4}, while for the charged black hole with a phantom Maxwell field~\eqref{eq64}, the Lloyd bound is violated. Of course, the result is not surprising because the phantom Maxwell field is a ghost field, and it violates the strong energy condition~\cite{Yang:2016awy}.

\section{Conclusion and discussion}
\label{Con}

Black hole physics is quite an interesting field, where gravity theory, quantum mechanics, thermodynamics, and statistical physics are entangled all together. Black hole thermodynamics reveals the holographic properties of gravity. The information loss paradox of black hole links black hole physics to information theory. Recent studies further expand the range, where it shows that black hole physics is also related to quantum information theory and quantum computation. In this paper we calculated the complexity growth of some holographic states dual to charged black holes with only one horizon. This calculation is based on the so-called Complexity-Action (CA) conjecture~\cite{Brown:2015bva,Brown:2015lvg}.

In this paper, in the spirit of CA duality, we calculated the complexity growth of some holographic states dual to charged black holes in AdS space. We paid special attention to the case of charged black holes with a single horizon. For these charged black holes, the inner horizon is absent and a spacelike singularity appears inside the event horizon. We considered three kinds of such black holes: one is the charged dilation black hole, where the inner horizon turns to be a spacelike singularity due to the existence of the dilaton field. The second one is the Born-Infeld black hole with $\beta^2 Q^2<1/4$. With this condition, the inner horizon is also absent. The third one is the charged phantom RN-AdS black hole, that is, the theory with a Maxwell field of negative kinetic term. The results are summarized as
\begin{align}
\hbox{charged dilation AdS BH}:&\quad\frac{\mathrm{d}S}{\mathrm{d}t}=2M-\mu_+Q-D,\quad D=\frac{Q^2\mathrm{e}^{2\phi_0}}{2M},\\
\hbox{charged Born-Infeld AdS BH}:&\quad\frac{\mathrm{d}S}{\mathrm{d}t}=2M-\mu_+Q-C,\quad C=\beta^\frac12Q^\frac32\frac{\Gamma(\frac14)\Gamma(\frac54)}{3\Gamma(\frac12)},\\
\hbox{charged phantom RN-AdS BH}:&\quad\frac{\mathrm{d}S}{\mathrm{d}t}=2M+\mu_+Q.
\end{align}

Following the approach proposed recently in~\cite{Lehner:2016vdi}, it was found that the action growth of the WDW patch for the first two kinds of charged black holes is always finite and well defined, and it satisfy the Lloyd bound. Namely although the asymptotic behavior of both black holes near the spacelike singularity is quite different from the one in the case of the AdS Schwarzschild black hole, the action growth turns out to be finite in both cases. In addition, we checked if the two different approaches proposed in~\cite{Lehner:2016vdi} and in~\cite{Brown:2015bva,Brown:2015lvg} give the same result. Although they are totally different from each other, we obtained the same result. For the charged black hole with a phantom Maxwell field, the action growth was also found to be finite after cancellation between the divergent part from the bulk contribution and the divergent part from the boundary contribution at the singularity. It should be noted, however, that the action growth in this case violates the Lloyd bound. This is not surprising since a phantom Maxwell field does not satisfy the strong energy condition~\cite{Yang:2016awy}.

The approach proposed by Lehner {\it et al.}~\cite{Lehner:2016vdi} follows from an earlier work by Hayward~\cite{Hayward:1993my} and defines a set of rules for all different joint terms, where the boundary contribution from null hypersurfaces is set to zero by affine parametrization. On the other hand, in the approach by Brown {\it et al.}~\cite{Brown:2015bva,Brown:2015lvg}, there is no contribution from joint terms; instead there does exist the boundary contribution from a spacelike/timelike surface that approaches the null hypersurface. Thus these two approaches look very different, and there seems to be no guarantee for them to give the same result. However, see \cite{Ruan:2017tkr} for a possible proof on this issue. These two independent approaches for calculating the gravitational action of spacetime with nonsmooth boundaries were previously compared within Einstein gravity~\cite{Lehner:2016vdi}. In the case of Gauss-Bonnet gravity, the action growth rate was computed in~\cite{Cai:2016xho} by utilizing the approach~\cite{Brown:2015bva,Brown:2015lvg}. However, the calculation by the approach~\cite{Lehner:2016vdi} is currently missing. It will be interesting to see if both approaches give the same result even for Gauss-Bonnet gravity, or more general theories of gravity.

Here, let us mention an alternative to the complexity=action conjecture~\cite{Brown:2015bva,Brown:2015lvg}; there are also some discussions
on the so-called ``complexity=volume'' conjectures in the literature. The first proposal of ``complexity=volume''~\cite{Stanford:2014jda} conjectured that the holographic description of complexity is proportional to the volume of the Einstein-Rosen (ER) bridge that connects two boundary states. However, the proportionality coefficient in this ``complexity=ER volume'' proposal cannot be uniquely determined, and the ER volume was in fact later found to be holographic dual to the information metric \cite{MIyaji:2015mia}. The second proposal of ``complexity=volume''~\cite{Alishahiha:2015rta,Barbon:2015ria,Barbon:2015soa} conjectured that the complexity is holographically dual to the volume of a time slice in the bulk enclosed by the minimal surface that appears in the calculation of holographic entanglement entropy~\cite{Ryu:2006bv}. See also~\cite{Momeni:2016ekm,Momeni:2016qfv} for this ``complexity=extremal volume'' proposal. In Ref.~\cite{Couch:2016exn} it was argued that the thermodynamical volume of a black hole may be identified with the late time action growth rate of the WDW patch. The thermodynamical volume of a black hole that appears in this ``complexity=BH volume'' proposal was also discussed in~\cite{Brown:2017jil} in which the uncomplexity is interpreted as the accessible volume of spacetime behind a black hole horizon.

Finally, let us stress that the action (volume) in the CA-duality (Complexity-Volume (CV)-duality) is in fact divergent in those proposals because of the infinite spacetime region extended all the way to the asymptotic boundary of the bulk geometry. To get a regularized action or volume, one may introduce  some cutoff surfaces near the  singularities and near asymptotic boundaries, compute the action or volume, and then subtract  corresponding contributions from  the vacuum AdS background ~\cite{Chapman:2016hwi,Carmi:2016wjl}.  However, the background subtraction approach suffers from some problems: one is that the resulting action depends on the choice of the reference background, and another is that sometimes no proper background could be chosen.  To get a finite and unique action, one more proper approach is to take the surface counterterm approach~\cite{Reynolds:2016rvl,Kim:2017lrw}, which is similar to the holographic renormalization technique developed in the AdS/CFT correspondence. This approach shares the following features: First, one needs not to specify a reference background like AdS vacuum; second, one also needs not to specify a special coordinate system in order to make a meaningful comparison between two spacetime geometries at the cutoff surfaces; third, the regularized action is independent of the reparametrizations of null generators and normalizations of null normal vectors; and last, the surface counterterms are determined alone by the boundary geometry. Thus a quite important and interesting issue is to calculate the complexity from the field theory side and to make a comparison with the resulting action in the CA conjecture.

\begin{acknowledgments}
RGC thanks the participants of JGRG 26 meeting for various discussions on relevant issues, and Li-Ming Cao, Shan-Ming Ruan, Ya-Wen Sun and Run-Qiu Yang for some useful discussions. This work was initialed during a visit of RGC as a visiting professor to the Yukawa Institute for Theoretical Physics, Kyoto University, the warm hospitality extended to him is greatly appreciated. RGC was supported in part by the  National Natural Science Foundation of China under Grants No.11375247, No.11435006 and No.11447601, and in part by a key project of CAS, Grant No.QYZDJ-SSW-SYS006. This work was also supported in part by the MEXT KAKENHI Nos.~15H05888 and 15K21733.
\end{acknowledgments}

\bibliographystyle{utphys}
\bibliography{ref}

\end{document}